\documentclass{elsart}
\usepackage{amssymb}
\usepackage{amsmath}
\usepackage{graphicx}
\usepackage[normalem]{ulem}
\usepackage[dvips]{color}
\renewcommand\sout{\bgroup \color{red} \ULdepth=-.5ex \ULset}

\begin{document}
\begin{frontmatter}
\title{Analytical relations between nuclear symmetry energy and single-nucleon potentials in isospin asymmetric nuclear matter}
\author [TAMUC,NJU]{Chang Xu},
\author [TAMUC]{Bao-An Li\thanksref{info}},
\author[TAMUC,SJTU]{Lie-Wen Chen},
\author[TAMU] {Che Ming Ko}
\address[TAMUC]{Department of Physics and Astronomy, Texas A$\&$M
University-Commerce, Commerce, Texas 75429-3011,
USA}
\address[NJU]{Department of Physics, Nanjing University, Nanjing
210008, China}
\address[SJTU]{Department of Physics, Shanghai Jiao Tong
University, Shanghai 200240, China}
\address[TAMU]{Cyclotron Institute and Department of Physics and Astronomy,
Texas A\&M University, College Station, TX 77843-3366, USA}
\thanks[info]{Corresponding author: Bao-An\_Li@Tamu-Commerce.edu}

\begin{abstract}
Using the Hugenholtz-Van Hove theorem, we derive general
expressions for the quadratic and quartic symmetry energies in
terms of single-nucleon potentials in isospin asymmetric nuclear
matter. These analytical relations are useful for gaining deeper
insights into the microscopic origins of the uncertainties in our
knowledge on nuclear symmetry energies especially at
supra-saturation densities. As examples, the formalism is applied
to two model single-nucleon potentials that are widely used in
transport model simulations of heavy-ion reactions.
\end{abstract}
\end{frontmatter}
\maketitle

\section{Introduction}
One of the central issues currently under intense investigation in
both nuclear physics and astrophysics is the Equation of State
(EOS) of neutron-rich nuclear matter \cite{JML04,AWS05,li1}. For
cold nuclear matter of isospin asymmetry
$\delta=(\rho_n-\rho_p)/(\rho_n+\rho_p)$ at density $\rho$, the
energy per nucleon $E(\rho,\delta)$ can be expressed as an even
series of $\delta$ that respects the charge symmetry of strong
interactions, namely,
$E(\rho,\delta)=E_0(\rho,0)+\sum_{i=2,4,6...}E_{sym,i}(\rho)\delta^i$
where $E_{sym,i}(\rho)$ is the so-called symmetry energy of the
\textit{i}th order \cite{li1} and $E_0(\rho,0)$ is the EOS of
symmetric nuclear matter. The quadratic term $E_{sym,2}(\rho)$ is
most important and its value at normal nuclear matter density
$\rho_0$ is known to be around 30 MeV from analyzing nuclear
masses within liquid-drop models. Essentially, all microscopic
many-body calculations have indicated that the higher-order terms
are usually negligible around $\rho_0$, leading to the so-called
empirical parabolic law of EOS even for $\delta$ approaching unity
for pure neutron matter. The $E_{sym,2}(\rho)$ is then generally
regarded as the symmetry energy. For instance, the value of the
quartic term has been estimated to be less than 1 MeV at $\rho_0$
\cite{sie70,lee98}. However, the presence of higher-order terms at
supra-saturation densities can significantly modify the proton
fraction in neutron stars at $\beta$-equilibrium and thus the
cooling mechanism of proto-neutron stars \cite{ste,zfs}. It was
also found that a tiny quartic term can cause a big change in the
calculated core-crust transition density in neutron stars
\cite{sjo,XCLM09}. Therefore, precise evaluations of the quartic
symmetry energy in neutron-rich matter are useful. Although much
information about the EOS of symmetric nuclear matter
$E_0(\rho,0)$ has been accumulated over the past four decades, our
knowledge about the density dependence of $E_{sym,i}(\rho)$ is
unfortunately still very poor. It has been generally recognized
that the $E_{sym,i}(\rho)$, especially the quadratic and quartic
terms, is critical for understanding not only the structure of
rare isotopes and the reaction mechanism of heavy-ion collisions,
but also many interesting issues in astrophysics
\cite{XCLM09,li0,bro,li2,dan,bar,Sum94,Bom01,LWC05,tsa,Cen09,Joe10,xia,wen}.
Therefore, to determine the $E_{sym,i}(\rho)$ in neutron-rich
matter has recently become a major goal in both nuclear physics
and astrophysics. While significant progress has been made
recently in constraining the $E_{sym,2}(\rho)$ especially around
and below the saturation density, see, e.g.,
\cite{LWC05,tsa,Cen09,Joe10}, much more work needs to be done to
constrain more tightly the $E_{sym,i}(\rho)$ at supra-saturation
densities where model predictions are rather diverse
\cite{Das03,ulr,van,zuo,Fri05,she,Che05c,zhli,Sto03,pan,wir,kut}.
As dedicated experiments using advanced new detectors have now
been planned to investigate the high density behavior of
$E_{sym,2}(\rho)$ at many radioactive beam facilities around the
world, it has become an urgent task to investigate theoretically
more deeply the fundamental origin of the extremely uncertain high
density behavior of $E_{sym,2}(\rho)$. It is also of great
interest to evaluate possible corrections due to the
$E_{sym,4}(\rho)$ term to the equation of state of asymmetric
nuclear matter.

Among existing proposals for extracting information about
$E_{sym,i}(\rho)$ using terrestrial laboratory experiments,
transport model simulations have shown that many observables in
heavy-ion reactions are particularly useful for studying
$E_{sym,i}(\rho)$ in a broad density range. In these transport
model simulations of heavy-ion reactions, the EOS enters the
reaction dynamics and affects the final observables through the
single-nucleon potential $U_{n/p}(\rho,\delta,k)$ where $k$ is the
nucleon momentum. Except in situations where statistical
equilibrium is established and thus many observables are directly
related to the binding energy $E(\rho,\delta)$ after correcting
for finite-size effects, what is being directly probed in
heavy-ion reactions is the single-nucleon potential
$U_{n/p}(\rho,\delta,k)$. The latter is, however, directly related
to the symmetry energy $E_{sym,2}(\rho)$ through the underlying
nuclear effective interaction as first pointed out by Brueckner,
Dabrowski and Haensel \cite{bru64,Dab73} using K-matrices within
the Brueckner theory. They showed that if one expands
$U_{n/p}(\rho,\delta,k)$ to the leading order in $\delta$ as in
the well-known Lane potential \cite{Lan62}, i.e.,
\begin{equation}\label{Lane}
U_{n/p}(\rho,\delta,k)\approx U_0(\rho,k) \pm
U_{sym,1}(\rho,k)\delta
\end{equation}
where $U_0(\rho,k)$ and $U_{sym,1}(\rho,k)$ are, respectively, the
nucleon isoscalar and isovector (symmetry) potentials, the
quadratic symmetry energy is then \cite{bru64,Dab73}
\begin{equation}\label{Esym}
E_{sym,2}(\rho)=\frac{1}{3} t(k_F) + \frac{1}{6} \frac{\partial
U_0}{\partial k}\mid _{k_F}\cdot k_F +
\frac{1}{2}U_{sym,1}(\rho,k_F)
\end{equation}
where $t(k_F)$ is the nucleon kinetic energy at the Fermi momentum
$k_F=(3\pi^2\rho/2)^{1/3}$ in symmetric nuclear matter of density
$\rho$. The above equation indicates that the symmetry energy
$E_{sym,2}(\rho)$ depends only on the single-particle kinetic and
potential energies at the Fermi momentum $k_F$. This is not
surprising since the microscopic origin of the symmetry energy is
the difference in the Fermi surfaces of neutrons and protons. The
first term $E_{sym}^{kin}=\frac{1}{3} t(k_F)=\frac{\hbar ^2}{6m}
(\frac{3\pi^2}{2})^{\frac{2}{3}} \rho^{\frac{2}{3}}$ is the
trivial kinetic contribution due to the different Fermi momenta of
neutrons and protons; the second term $\frac{1}{6} \frac{\partial
U_0}{\partial k}\mid _{k_F}\cdot k_F$ is due to the momentum
dependence of the isoscalar potential and also the fact that
neutrons and protons have different Fermi momenta; while the term
$\frac{1}{2}U_{sym}(\rho,k_F)$ is due to the explicit isospin
dependence of the nuclear strong interaction. For the isoscalar
potential $U_0(\rho,k)$, reliable information about its density
and momentum dependence has already been obtained from high energy
heavy-ion collisions, see, e.g., ref. \cite{dan}, albeit there are
still some rooms for further improvements, particularly at high
momenta/densities. On the contrary, the isovector potential
$U_{sym,1}(\rho,k)$ is still not very well determined, especially
at high densities and momenta, and has been identified as the key
quantity responsible for the uncertain high density behavior of
the symmetry energy as stressed in ref.\cite{li1}.

In the present work, we first show, using both the differential
and integral formulations of the Hugenholtz-Van Hove (HVH) theorem
\cite{hug}, that the relation in Eq.(\ref{Esym}) is valid in
general. We then derive an expression for the quartic symmetry
energy $E_{sym,4}(\rho)$ in terms of the single-nucleon potential
by keeping higher-order terms in the expansion of both the EOS and
the single-nucleon potential. Applying the HVH formalism to two
model single-nucleon potentials, namely, the
Bombaci-Gale-Bertsch-Das Gupta (BGBD) potential \cite{Bom01} and a
modified Gogny Momentum-Dependent-Interaction
(MDI)\cite{Das03,dec}, which are among the most widely used ones
in studying isospin physics based on transport model simulations
of heavy-ion reactions \cite{li1,bar}, we examine the relative
contributions from the kinetic and various potential terms to
$E_{sym,2}(\rho)$ and $E_{sym,4}(\rho)$. We put the emphasis on
identifying those terms that dominate the high density behaviors
of $E_{sym,2}(\rho)$. Finally, we evaluate the relative importance
of the $E_{sym,4}(\rho)$ term by studying the $E_{sym,4}(\rho)/
E_{sym,2}(\rho)$ ratio as a function of density.

The paper is organized as follows. In Section~\ref{symmetry},
based on the HVH theorem we derive general expressions for the
higher-order symmetry energy terms $E_{sym,2}(\rho)$ and
$E_{sym,4}(\rho)$ in terms of the single-nucleon isoscalar and
isovector potentials. The derivation is carried out in
Section~\ref{hvh} using the differential form of the HVH theorem
by starting from the neutron and proton chemical potentials and in
Section~\ref{fermi} using the integral form of the HVH theorem by
starting from the total energy density of the system. Numerical
results and discussions for both the BGBD and MDI interactions are
given in Sections~\ref{BGBD} and \ref{MDI}, respectively. Finally,
we give a summary in Section~\ref{summary}.

\section{Symmetry energy in terms of the single-nucleon potential}\label{symmetry}

In this section, we present two alternative approaches to derive
expressions for the quadratic and quartic symmetry energy terms
$E_{sym,2}(\rho)$ and $E_{sym,4}(\rho)$. Both are based on the
Fermi gas model of interacting nucleons and satisfy the HVH
theorem that was first derived in Ref.~\cite{hug}. There are,
however, some technical differences between the two approaches in
that a Taylor-series expansion is made on the single-nucleon
energy in one approach but on the total energy of the system in
the other.

\subsection{Derivation using the Hugenholtz-Van Hove theorem}\label{hvh}

The Hugenholtz-Van Hove theorem \cite{hug} describes a fundamental
relation among the Fermi energy $E_{F}$, the average energy per
particle $E$ and the pressure of the system $P$ at the absolute
temperature of zero. For a one-component system, in terms of the
energy density $\xi=\rho E$, the general HVH theorem can be
written as \cite{hug,Sat99}
\begin{eqnarray}\label{HVH}
E_{F}=\frac{d\xi}{d \rho}=\frac{d (\rho E)}{d \rho} =  E+ \rho
\frac{d E}{d \rho} = E+ P/\rho.
\end{eqnarray}
The above relation has been strictly proven to be valid for any
interacting self-bound infinite Fermi system. It does not depend
upon the precise nature of the interaction. In the special case of
nuclear matter at saturation density where the pressure $P$
vanishes, the average energy per nucleon becomes equal to the
Fermi energy, i.e., $E_{F}=E$. It is worthwhile to stress that the
general HVH theorem of Eq.(\ref{HVH}) is valid at any arbitrary
density as long as the temperature remains zero \cite{hug,Sat99}.
In fact, a successful theory for nuclear matter is required not
only to describe satisfactorily all saturation properties of
nuclear matter but also to fulfill the general HVH theorem at any
density. In the following, we use the general HVH theorem to
derive the relation between the nuclear symmetry energy and the
single-nucleon potential.

According to the HVH theorem, the chemical potentials of neutrons
and protons in isospin asymmetric nuclear matter of energy density
$\xi(\rho,\delta)=\rho E(\rho,\delta)$ are, respectively
\cite{hug,Sat99},
\begin{eqnarray} t(k_F^n)+U_n(\rho,\delta,k_F^n) =
\frac{\partial \xi }{\partial \rho_n}, \label{chemUn}
\\
t(k_F^p)+U_p(\rho,\delta,k_F^p) = \frac{\partial \xi }{\partial
\rho_p}, \label{chemUp}
\end{eqnarray}
where $t(k)=\hbar k^2/2m$ is the kinetic energy and $U_{n/p}$ is
the neutron/proton single-particle potential. The Fermi momenta
of neutrons and protons are $k_F^n=k_F(1+\delta)^{1/3}$ and
$k_F^p=k_F(1-\delta)^{1/3}$, respectively. Subtracting
Eq.(\ref{chemUp}) from Eq.(\ref{chemUn}) gives \cite{bru64,Dab73}
\begin{eqnarray}\label{UnminusUp}
[t(k_F^n)-t(k_F^p)]+[U_n(\rho,\delta,k_F^n)-U_p(\rho,\delta,k_F^p)]=\frac{\partial
\xi }{\partial \rho_n}-\frac{\partial \xi }{\partial \rho_p}.
\end{eqnarray}
The nucleon single-particle potentials can be expanded as a power
series of $\delta$ while respecting the charge symmetry of nuclear
interactions under the exchange of neutrons and protons,
\begin{eqnarray}\label{UnUp1}
U_n(\rho,\delta,k) & = & U_0(\rho,k) + \sum_{i=1,2,3...}U_{sym,i}(\rho,k) \delta^i  \nonumber\\
& = & U_0(\rho,k) + U_{sym,1}(\rho,k) \delta +U_{sym,2}(k) \delta^2 +...
\end{eqnarray}
\begin{eqnarray}\label{UnUp2}
U_p(\rho,\delta,k) & = & U_0(\rho,k) + \sum_{i=1,2,3...}U_{sym,i}(\rho,k) (-\delta)^i  \nonumber\\
& = & U_0(\rho,k) - U_{sym,1}(\rho,k) \delta +U_{sym,2}(\rho,k) \delta^2 -....
\end{eqnarray}
If one neglects the higher-order terms ($\delta^2$,
$\delta^3$,...), Eq.(\ref{UnUp1}) and Eq.(\ref{UnUp2}) reduce to
the Lane potential in Eq.(\ref{Lane}). Expanding both the kinetic
and potential energies around the Fermi momentum $k_F$, the left
side of Eq.(\ref{UnminusUp}) can be further written as
\begin{eqnarray}\label{leftside}
&&[t(k_F^n)-t(k_F^p)]+[U_n(\rho,\delta,k_F^n)-U_p(\rho,\delta,k_F^p)] \nonumber\\
=&&\sum_{i=1,2,3...}\frac{1}{i!}\frac{\partial^i [t(k)+U_0(\rho,k)]}{\partial k^i}|_{k_F} k_F^i \nonumber\\
\times&&[(\sum\limits_{j=1,2,3..}F(j)\delta^j)^i-(\sum\limits_{j=1,2,3..}F(j)(-\delta)^j)^i] \nonumber\\
+ && \sum_{l=1,2,3...}U_{sym,l}(\rho,k_F)[\delta^{l}-(-\delta)^{l}] \nonumber\\
+ &&\sum_{l=1,2,3...}\sum_{i=1,2,3...}\frac{1}{i!}\frac{\partial^iU_{sym,l}(\rho,k)}{\partial k^i}|_{k_F} k_F^i
\nonumber \\
\times &&  [(\sum\limits_{j=1,2,3..}F(j)\delta^j)^i \delta^{l}-(\sum\limits_{j=1,2,3..}F(j)(-\delta)^j)^i (-\delta)^{l}]
\nonumber \\
= && [\frac{2}{3} \frac{\partial
[t(k)+U_0(\rho,k)]}{\partial k}|_{k_F}k_F + 2 U_{sym,1}(\rho,k_F)]\delta + ...,
\end{eqnarray}
where we have introduced the function
$F(j)=\frac{1}{j!}[\frac{1}{3}(\frac{1}{3}-1)...(\frac{1}{3}-j+1)]$.
For the right side of Eq.(\ref{UnminusUp}), expanding in powers of
$\delta$ gives
\begin{eqnarray} \label{rightside}
&& \frac{\partial \xi }{\partial \rho_n}-\frac{\partial \xi
}{\partial \rho_p}= \frac{2}{\rho} \frac{\partial \xi}{\partial
\delta} = \sum_{i=2,4,6...}2i E_{sym,i}(\rho)\delta^{i-1} \nonumber \\
&& =4E_{sym,2}(\rho) \delta + 8E_{sym,4}(\rho) \delta^3 +12E_{sym,6}(\rho) \delta^5+....
\end{eqnarray}
Comparing the coefficient of each $\delta^i$ term in
Eq.(\ref{leftside}) with that in Eq.(\ref{rightside}) then gives
the symmetry energy of any order. For instance, the quadratic term
\begin{eqnarray}\label{Esym2}
E_{sym,2}(\rho) &=& \frac{1}{6} \frac{\partial
[t(k)+U_0(\rho,k)]}{\partial k}|_{k_F}k_F + \frac{1}{2}
U_{sym,1}(\rho,k_F)\nonumber \\ &=&\frac{1}{3} t(k_F) + \frac{1}{6}
\frac{\partial U_0}{\partial k}\mid _{k_F}\cdot k_F +
\frac{1}{2}U_{sym,1}(\rho,k_F)
\end{eqnarray}
is identical to that in Eq.(\ref{Esym}), while the quartic term
can be written as
\begin{eqnarray}\label{Esym4}
&& E_{sym,4}(\rho) = \left[ \frac{5}{324}\frac{\partial
[t(k)+U_0(\rho,k)]}{\partial k}|_{k_F} k_F \right.
\nonumber \\ && \left. - \frac{1}{108}
\frac{\partial^2 [t(k)+U_0(\rho,k)]}{\partial k^2}|_{k_F} k_F^2
+\frac{1}{648} \frac{\partial^3 [t(k)+U_0(\rho,k)]}{\partial
k^3}|_{k_F} k_F^3 \right.\nonumber \\&&
\left. - \frac{1}{36} \frac{ \partial U_{sym,1}(\rho,k)}{\partial
k}|_{k_F} k_F + \frac{1}{72} \frac{ \partial^2
U_{sym,1}(\rho,k)}{\partial k^2}|_{k_F} k_F^2 \right.
\nonumber \\&& \left. + \frac{1}{12}
\frac{\partial U_{sym,2}(\rho,k)}{\partial k}|_{k_F} k_F +
\frac{1}{4} U_{sym,3}(\rho,k_F)\right].
\end{eqnarray}

\subsection{Derivation using the total energy of an interacting Fermi gas}\label{fermi}

The symmetry energy of any order obtained in the previous
subsection can also be derived within the interacting Fermi gas
model \cite{pre,xuli}. Although the derivation is more tedious, it
is physically interesting and mathematically instructional.

There are different ways to calculate the total energy of a Fermi
system at a given density $\rho$. As explained in detail by
Bertsch and Das Gupta \cite{BD88}, one could determine the total
energy starting with empty space, adding particles until the
desired density is reached. Each added particle would contribute
an energy $k(\rho_x)^2/2m+U(\rho_x,k(\rho_x))$, where $k$ is the
Fermi momentum corresponding to the density $\rho_x$ of particles
already added to the system. As stressed by Bertsch and Das Gupta,
the single-particle potential $U(\rho_x,k(\rho_x))$ is not the
same as the potential energy per particle as one might at first
guess. In this way, the total energy density of the asymmetric
nuclear matter written in coordinate space is
\begin{equation}\label{erspace}
\xi=\int_0^{\rho^n} [k(\rho_x)^2/2m+U_n(\rho_x,k(\rho_x))] d\rho_x
+ \int_0^{\rho^p} [k(\rho_x)^2/2m+U_p(\rho_x,k(\rho_x))]d\rho_x,
\end{equation}
where $\rho^n+\rho^p=\rho$. The single particle potential
$U(\rho_x,k(\rho_x))$ for the neutron or proton in the first or
second integral in Eq.(\ref{erspace}) can be rewritten as
\begin{eqnarray}
&&U(\rho_x,k(\rho_x))=U(\rho,\delta^*,k(\rho_x)),
\end{eqnarray}
where $\rho_x=\rho(1+\delta^*)/2$ in terms of the local isospin
asymmetry $\delta^*$ (to be discussed in detail in the following).
The single particle potential can be further expanded as a series
of $\delta^*$
\begin{eqnarray}
&&U(\rho,\delta^*,k(\rho_x))=U_0(\rho,k(\rho_x))+\sum_{i=1,2,3...}U_{sym,i}(\rho,k(\rho_x))(\delta^*)^i
\\ \nonumber
&&=U_0(\rho,k(\rho_x))+U_{sym,1}(\rho,k(\rho_x))(\delta^*)+U_{sym,2}(\rho,k(\rho_x))(\delta^*)^2+....
\end{eqnarray}
Similar to the interpretation of Eq.(\ref{erspace}), one can raise
the momenta of all particles in the system from zero to the Fermi
momentum corresponding to the density $\rho$. In either the
coordinate or momentum space, as one increases the density or
momentum for both neutrons and protons to build up the desired
Fermi system, the local isospin asymmetry $\delta^*$ changes
continuously as a function of density or momentum, i.e.,
\begin{eqnarray}
\delta^*=\left[\frac{2\rho_x}{\rho}-1\right]\rightarrow
\delta^*=\left[\frac{k^3}{k_F^3}-1\right]
\end{eqnarray}
where $\rho_x$ is either the neutron or proton density used in the
two terms of Eq.(\ref{erspace}), and $k$ is the neutron or proton
momentum used in the two terms of Eq.(\ref{ekspace}) below. At the
respective Fermi surfaces of neutrons and protons, the local
isospin asymmetry $\delta^*$ reduce to the global one
\begin{eqnarray}
&&\delta^*(k_F^n)=\left[\frac{2\rho^n}{\rho}-1\right]=\left[\frac{(k_F^n)^3}{k_F^3}-1\right]=
\delta, \nonumber\\
&&\delta^*(k_F^p)=\left[\frac{2\rho^p}{\rho}-1\right]=\left[\frac{(k_F^p)^3}{k_F^3}-1\right]=
(-\delta).
\end{eqnarray}

In order to make use of the same technique in expanding both the
kinetic and potential terms in $\delta$ series, it is more
convenient to express the energy density in momentum space
\begin{equation}\label{ekspace}
\xi=\frac{1}{\pi^2} \left[\int_0^{k_F^n}
[t(k)+U_n(\rho,\delta^*,k)] k^2dk + \int_0^{k_F^p}
[t(k)+U_p(\rho,\delta^*,k)] k^2dk\right].
\end{equation}
Before we proceed to derive expressions for $E_{sym,2}(\rho)$ and
$E_{sym,4}(\rho)$, it is critical to examine whether the energy
density given in Eq.(\ref{erspace}) or Eq.(\ref{ekspace})
satisfies the HVH theorem. Noticing that
\begin{eqnarray}
\frac{\partial \xi }{\partial \rho^n} &=& \frac{\partial \xi
}{\partial k_F^n} \frac{\partial k_F^n}{\partial \rho^n}=
\frac{\partial \xi }{\partial k_F^n} /\left (\frac{\partial
\rho^n}{\partial k_F^n}\right) = \frac{\partial \xi }{\partial k_F^n} /
\left(\frac{\partial [\frac{(k_F^n)^3}{3\pi^2}]}{\partial k_F^n}\right) =
\frac{\partial \xi }{\partial k_F^n} \left[\frac{\pi^2}{(k_F^n)^2}\right],
\end{eqnarray}
then it is straightforward to show using Eq.(\ref{ekspace}) that
\begin{eqnarray}
\frac{\partial \xi }{\partial \rho^n} &=& \left\{\frac{1}{\pi^2}
[t(k_F^n)+U_n(\rho,\delta,k_F^n)] (k_F^n)^2 \right\}
\left[\frac{\pi^2}{(k_F^n)^2}\right]
\nonumber\\
&=& t(k_F^n)+U_n(\rho,\delta,k_F^n).
\end{eqnarray}
Similarly, one can show that
\begin{eqnarray}
\frac{\partial \xi }{\partial \rho^p}  =
t(k_F^p)+U_n(\rho,\delta,k_F^p).
\end{eqnarray}
Thus, the HVH theorem is indeed satisfied by the energy density
expressed in both Eq.(\ref{erspace}) and Eq.(\ref{ekspace}).

To obtain expressions for $E_{sym,2}(\rho)$ and $E_{sym,4}(\rho)$,
we make a Taylor-series expansion of the energy density in
Eq.(\ref{ekspace}). To proceed, it is useful to first recall that
for any continuous function $Y(k)$ the Taylor-series expansion of
the integral $ f(k_F^i)=\int_0^{k_F^i} Y(k) k^2 dk $ around $k_F$
leads to \cite{pre}
\begin{eqnarray}\label{ts}
f(k_F^i)&=&f(k_F) + Y(k_F)\cdot k_F^2\cdot (k_F^i-k_F)
\nonumber \\&+& \frac{1}{2} \left[\frac{\partial Y}{\partial
k_F^i}|_{k_F}\cdot k_F^2 + 2k_F\cdot Y(k_F)\right]\cdot(k_F^i-k_F)^2+...,
\end{eqnarray}
Moreover, it is easy to show that
\begin{eqnarray}
k_F^n-k_F
 &  = & \left[\frac{1}{3}\delta-\frac{1}{9}\delta^2+\frac{5}{81}\delta^3-\frac{10}{243}\delta^4
+ ...\right]k_F \nonumber \\k_F^p-k_F
 &  = & \left[-\frac{1}{3}\delta-\frac{1}{9}\delta^2-\frac{5}{81}\delta^3-\frac{10}{243}\delta^4+
... \right]k_F.
\end{eqnarray}

The kinetic energy per nucleon can then be expanded as
\begin{eqnarray}\label{kinetic}
\overline{T} &  = & \alpha \int_0^{k_F^n} t(k) k^2dk + \alpha
\int_0^{k_F^p} t(k) k^2dk
\nonumber \\&  = &  2\alpha \int_0^{k_F} t(k) k^2dk
 + \alpha t(k_F) k_F^2 [(k_F^n-k_F)+(k_F^p-k_F)]
\nonumber \\& + &  \frac{\alpha}{2!} \frac{\partial [t(k)
k^2]}{\partial k}|_{k_F}[(k_F^n-k_F)^2+(k_F^p-k_F)^2]
\nonumber \\& + & \frac{\alpha}{3!} \frac{\partial^2 [t(k)
k^2]}{\partial k^2}|_{k_F} [(k_F^n-k_F)^3+(k_F^p-k_F)^3]
\nonumber \\& + & \frac{\alpha}{4!} \frac{\partial^3 [t(k)
k^2]}{\partial k^3}|_{k_F}[(k_F^n-k_F)^4+(k_F^p-k_F)^4]
\nonumber \\&  = &
 2 \alpha \int_0^{k_F} t(k) k^2dk + \frac{1}{6}
\frac{\partial t(k)}{\partial k}\mid_{k_F}k_F \delta^2
\nonumber \\& + & \left[\frac{5}{324}\frac{\partial t(k)}{\partial
k}\mid_{k_F}k_F - \frac{1}{108}\frac{\partial^2 t(k)}{\partial
k^2}\mid_{k_F}k_F^2 +\frac{1}{648}\frac{\partial^3 t(k)}{\partial
k^3}\mid_{k_F}k_F^3 \right]\delta^4 + ...,\nonumber\\
\end{eqnarray}
where $\alpha=3 / (2k_F^3)$. Thus, the kinetic contribution to the
$E_{sym,2}(\rho)$ is
\begin{eqnarray}
E_{sym,2}^{kin}(\rho)=\frac{1}{6} \frac{\partial t(k)}{\partial
k}|_{k_F} k_F = \frac{1}{3} t(k_F) = \frac{\hbar^2}{6m}
\left(\frac{3\pi^2}{2}\right)^{2/3} \rho^{2/3},
\end{eqnarray}
and its contribution to  $E_{sym,4}(\rho)$ is
\begin{eqnarray}
E_{sym,4}^{kin}(\rho) & = & \frac{5}{324}\frac{\partial
t(k)}{\partial k}|_{k_F} k_F - \frac{1}{108} \frac{\partial^2
t(k)}{\partial k^2}|_{k_F} k_F^2 +\frac{1}{648} \frac{\partial^3
t(k)}{\partial k^3}|_{k_F} k_F^3
\nonumber \\& = &  \frac{1}{81} t(k_F) = \frac{\hbar^2}{162m}
\left(\frac{3\pi^2}{2}\right)^{2/3} \rho^{2/3}.
\end{eqnarray}

For the potential energy per nucleon written in momentum space, we
first expand the single-nucleon potential $U_{n/p}$ in local
isospin asymmetry $\delta^*(k)$
\begin{eqnarray}
\overline{U}  & = &\alpha \int_0^{k_F^n} U_n(\rho,\delta^*,k)k^2dk
+ \alpha \int_0^{k_F^p} U_p(\rho,\delta^*,k) k^2dk
\nonumber \\& = & [\alpha \int_0^{k_F^n} U_0(\rho,k) k^2dk + \alpha
\int_0^{k_F^p} U_0(\rho,k) k^2dk]
\nonumber \\&+& \sum_{i=1,2,3...}\left[\alpha \int_0^{k_F^n}
U_{sym,i}(\rho,k)(\delta^*)^i k^2dk + \alpha \int_0^{k_F^p}
U_{sym,i}(\rho,k)(\delta^*)^i k^2dk\right].\nonumber\\
\end{eqnarray}
We then expand each integral in the above equation around the
Fermi momentum $k_F$ using Eq.~(\ref{ts}). The $U_0$ contribution
to $\overline{U}$ is
\begin{eqnarray}
&& E_0^{pot}(\rho)=\alpha \int_0^{k_F^n} U_0(\rho,k)k^2dk + \alpha
\int_0^{k_F^p} U_0(\rho,k) k^2dk
\nonumber \\&=&  2 \alpha \int_0^{k_F} U_0(\rho,k) k^2dk +
\frac{1}{6} \frac{\partial U_0(\rho,k)}{\partial k}|_{k_F} k_F
\delta^2
\nonumber \\&+& \left[\frac{5}{324}\frac{\partial
U_0(\rho,k)}{\partial k}|_{k_F} k_F - \frac{1}{108}
\frac{\partial^2 U_0(\rho,k)}{\partial k^2}|_{k_F} k_F^2
+\frac{1}{648} \frac{\partial^3 U_0(\rho,k)}{\partial k^3}|_{k_F}
k_F^3 \right] \delta^4+....\nonumber\\
\end{eqnarray}
Contributions from the higher-order potential terms in $\delta$
can be obtained similarly. For the second order-symmetry energy
$E_{sym,2}(\rho)$, only the $U_{sym,1}$ term contributes
\begin{eqnarray}
&& E_1^{pot}(\rho)=\alpha \int_0^{k_F^n} U_{sym,1}(\rho,k)\delta^*
k^2dk + \alpha \int_0^{k_F^p} U_{sym,1}(\rho,k)\delta^* k^2dk
\nonumber \\& = &  \alpha \int_0^{k_F^n} U_{sym,1}(\rho,k) [(\frac{k}{k_F})^3-1] k^2dk
+ \alpha \int_0^{k_F^p} U_{sym,1}(\rho,k) [(\frac{k}{k_F})^3-1]
k^2dk\nonumber \\& = & 2\alpha \int_0^{k_F} U_{sym,1}(\rho,k) [(\frac{k}{k_F})^3-1] k^2dk
\nonumber \\&+& \frac{1}{2} U_{sym,1}(\rho,k_F) \delta^2 + \left[ -
\frac{1}{36} \frac{
\partial U_{sym,1}(\rho,k)}{\partial k}|_{k_F} k_F + \frac{1}{72} \frac{
\partial^2 U_{sym,1}(\rho,k)}{\partial k^2}|_{k_F} k_F^2 \right]
\delta^4+....\nonumber\\
\end{eqnarray}
To obtain the fourth-order symmetry energy $E_{sym,4}(\rho)$ , all
$U_{sym,1}$, $U_{sym,2}$, and $U_{sym,3}$ terms are needed, i.e.
\begin{eqnarray}
&& E_2^{pot}(\rho)=\alpha \int_0^{k_F^n}
U_{sym,2}(\rho,k)(\delta^*)^2 k^2dk + \alpha \int_0^{k_F^p}
U_{sym,2}(\rho,k)(\delta^*)^2 k^2dk
\nonumber \\& = & \alpha \int_0^{k_F^n} U_{sym,2}(\rho,k)\left [\left(\frac{k}{k_F}\right)^3-1\right]^2 k^2dk
+ \alpha \int_0^{k_F^p} U_{sym,2}(\rho,k) [(\frac{k}{k_F})^3-1]^2
k^2dk\nonumber \\& = & 2\alpha \int_0^{k_F} U_{sym,2}(\rho,k)
\left[\left(\frac{k}{k_F}\right)^3-1\right]^2 k^2dk + \frac{1}{12} \frac{\partial
U_{sym,2}(\rho,k)}{\partial k}|_{k_F} k_F \delta^4+...,\nonumber\\
\end{eqnarray}
and
\begin{eqnarray}
&&E_3^{pot}(\rho)=\alpha \int_0^{k_F^n}
U_{sym,3}(\rho,k)(\delta^*)^3 k^2dk + \alpha \int_0^{k_F^p}
U_{sym,3}(\rho,k)(\delta^*)^3 k^2dk
\nonumber \\& = & \alpha \int_0^{k_F^n} U_{sym,3}(\rho,k) \left[\left(\frac{k}{k_F}\right)^3-1\right]^3 k^2dk
+ \alpha \int_0^{k_F^p} U_{sym,3}(\rho,k) [(\frac{k}{k_F})^3-1]^3
k^2dk\nonumber \\& = & 2\alpha \int_0^{k_F} U_{sym,3}(\rho,k)
\left[\left(\frac{k}{k_F}\right)^3-1\right]^3 k^2dk + \frac{1}{4} U_{sym,3}(\rho,k_F)
\delta^4+....\nonumber\\
\end{eqnarray}
Combining all coefficients of the $\delta^2$ and $\delta^4$ terms
in both the kinetic and potential parts given above, one sees that
the $E_{sym,2}(\rho)$ and $E_{sym,4}(\rho)$ obtained here are
exactly the same as those given in using directly the HVH theorem.
Moreover, they agree with earlier results obtained by Brueckner
{\it et al}. \cite{bru64,Dab73}.

\section{Applications and discussions}

As shown in the previous section, the symmetry energy can be
explicitly separated into the kinetic energy term $T$ and the
potential terms $U_0$ and $U_{sym,i}$ at the Fermi momentum $k_F$.
To evaluate their relative contributions to the symmetry energies,
especially for the second-order and fourth-order terms
$E_{sym,2}(\rho)$ and $E_{sym,4}(\rho)$, we consider in this
section two typical single-nucleon potentials that have been
widely used in tansport model simulations of heavy-ion reactions.

\subsection{The Bombaci-Gale-Bertsch-Das Gupta potential}\label{BGBD}

As a first example, we use the phenomenological potential of
Bombaci-Gale-Bertsch-Das Gupta \cite{Bom01}
\begin{eqnarray} \label{Bomba}
&&U_\tau(u,\delta,k)= Au+Bu^\sigma
-\frac{2}{3}(\sigma-1)\frac{B}{\sigma+1} (\frac{1}{2}+x_{3})
u^{\sigma}\delta^2
\nonumber\\
&&\pm \left[{-\frac{2}{3}A (\frac{1}{2}+x_{0}) u -
\frac{4}{3}\frac{B}{\sigma+1}(\frac{1}{2}+x_{3})u^{\sigma}\,}\right]\delta\nonumber\\
&& +\frac{4}{5\rho_0}\left[{\frac{1}{2} (3C-4z_1) \mathcal{I_{\tau}}
+ (C+2z_1)\mathcal{{I_{\tau^{\prime}}}}}\right]+ \left({C \pm
\frac{C-8z_1}{5}\delta}\right)u\cdot g(k),
\end{eqnarray}
where $u=\rho/\rho_0$ is the reduced density and $\pm$ is for
neutrons/protons. In the above, we have
$\mathcal{I}_\tau=[2/(2\pi)^3]\int d^3k f_{\tau}(k)g(k)$ with
$g(k)= 1/[{1+({\frac{k}{\Lambda}})^2 }]$ being a momentum
regulator and $f_{\tau}(k)$ being the phase space distribution
function. The parameter $\Lambda$ has the value
$\Lambda=1.5k_F^0$, where $k_F^0$ is the nucleon Fermi wave number
in symmetric nuclear matter at $\rho_0$. With A=-144 MeV, B=203.3
MeV, C=-75 MeV and $\sigma=7/6$, the BGBD potential reproduces all
ground state properties including an incompressibility $K_0$=210
MeV for symmetric nuclear matter \cite{Bom01}. The three
parameters $x_0, x_3$ and $z_1$ can be adjusted to give different
symmetry energy $E_{sym,2}(\rho)$ and the neutron-proton effective
mass splitting $m^*_n-m^*_p$ \cite{Bom01,riz04,Li04}. For example,
the parameter set $z_1=-36.75$ MeV, $x_0=-1.477$ and $x_3=-1.01$
leads to $m_n^*>m_p^*$ while the one with $z_1=50$ MeV,
$x_0=1.589$ and $x_3=-0.195$ leads to $m_n^*<m_p^*$ at all
non-zero densities and isospin asymmetries.

On expanding the BGBD potential in $\delta$, the coefficients of
the first few terms are
\begin{eqnarray}
U_0(\rho,k)&=& U_{n/p}|\, _{\delta=0}
\nonumber \\&=&   Au+Bu^\sigma +\frac{2C}{\rho_0}
\frac{\Lambda^2}{\pi^2} \left[k_F - \Lambda
\textrm{tan}^{-1}\left(\frac{k_F}{\lambda}\right)\right]+ C u\cdot
g(k),
\end{eqnarray}
\begin{eqnarray}
U_{sym,1}(\rho,k)&=&\pm \frac{1}{1!} \frac{\partial
U_{n/p}}{\partial \delta}|\, _{\delta=0}\nonumber\\
&=& \left[{-\frac{2}{3}A (\frac{1}{2}+x_{0}) u -
\frac{4}{3}\frac{B}{\sigma+1}(\frac{1}{2}+x_{3})u^{\sigma}\,}\right]\nonumber\\
&+& \frac{C-8z_1}{5}u\cdot g(k_F) + \frac{C-8z_1}{5}u\cdot g(k),
\end{eqnarray}
\begin{eqnarray}
U_{sym,2}(\rho,k) &=& \frac{1}{2!} \frac{\partial^2
U_{n/p}}{\partial \delta^2}|\, _{\delta=0}
\nonumber \\&=& -\frac{2}{3}(\sigma-1)\frac{B}{\sigma+1}
(\frac{1}{2}+x_{3}) u^{\sigma} - \frac{C}{3}u
\frac{k_F^2}{\Lambda^2} g(k_F)^2,
\end{eqnarray}
\begin{eqnarray}
U_{sym,3}(\rho,k) &=&  \pm \frac{1}{3!} \frac{\partial^3
U_{n/p}}{\partial \delta^3}|\, _{\delta=0}
\nonumber \\&=& \frac{C-8z_1}{135}u \frac{k_F^2}{\Lambda^2}(\frac{5k_F^2}{\Lambda^2}+1)
g(k_F)^3.
\end{eqnarray}

Thus, the second-order symmetry energy $E_{sym,2}(\rho)$ is given
by
\begin{eqnarray}
E_{sym,2}(\rho)  &=& \frac{1}{3} t(k_F) + \frac{1}{6}
\frac{\partial U_0}{\partial k}\mid _{k_F}\cdot k_F +
\frac{1}{2}U_{sym,1}(\rho,k_F)
\nonumber\\ &=& \frac{\hbar^2}{6m}
\left(\frac{3\pi^2}{2}\right)^{2/3} \rho^{2/3} - \frac{C}{3}u
\frac{k_F^2}{\Lambda^2} g(k_F)^2 \nonumber \\&+&
\left[{-\frac{1}{3}A (\frac{1}{2}+x_{0}) u -
\frac{2}{3}\frac{B}{\sigma+1}(\frac{1}{2}+x_{3})u^{\sigma}\,}\right]
\nonumber\\&+& \frac{C-8z_1}{5}u\cdot g(k_F),
\end{eqnarray}
and the fourth-order symmetry energy $E_{sym,4}(\rho)$ is
\begin{eqnarray}
&&E_{sym,4}(\rho) = \frac{\hbar^2}{162m}
\left(\frac{3\pi^2}{2}\right)^{2/3} \rho^{2/3}
\nonumber \\& + &
\left[ \frac{5}{324}\frac{\partial U_0(\rho,k)}{\partial k}|_{k_F}
k_F - \frac{1}{108} \frac{\partial^2 U_0(\rho,k)}{\partial
k^2}|_{k_F} k_F^2 +\frac{1}{648} \frac{\partial^3
U_0(\rho,k)}{\partial k^3}|_{k_F} k_F^3 \right.
\nonumber \\& - & \left.
\frac{1}{36} \frac{ \partial U_{sym,1}(\rho,k)}{\partial k}|_{k_F}
k_F + \frac{1}{72} \frac{ \partial^2 U_{sym,1}(\rho,k)}{\partial
k^2}|_{k_F} k_F^2 + \frac{1}{12} \frac{\partial
U_{sym,2}(\rho,k)}{\partial k}|_{k_F} k_F + \frac{1}{4}
U_{sym,3}(\rho,k_F)\right]
\nonumber \\& =  &
\frac{\hbar ^2}{162m} (\frac{3\pi^2 \rho}{2})^{\frac{2}{3}}-
\frac{C}{81} u
\frac{k_F^2}{\Lambda^2}(\frac{10k_F^4}{\Lambda^4}+\frac{5k_F^2}{\Lambda^2}+1)
g(k_F)^4+ \frac{C-8z_1}{135}u
\frac{k_F^2}{\Lambda^2}(\frac{5k_F^2}{\Lambda^2}+1) g(k_F)^3.\nonumber\\
\end{eqnarray}

\begin{figure}[!htb]
\centering
\includegraphics[width=12cm]{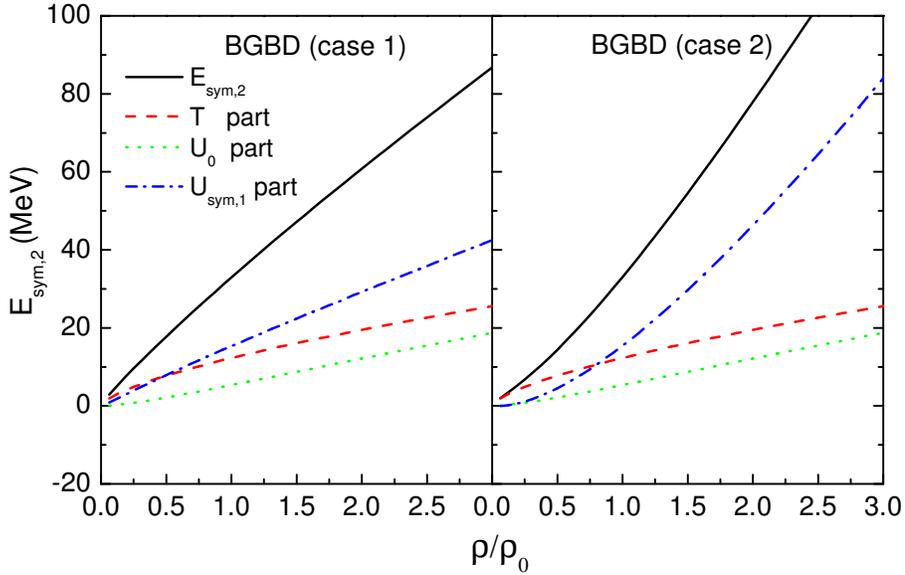}
\caption{The kinetic energy part (T), the isoscalar potential part
($U_0$) and the isovector potential part ($U_{sym,1}$) of the
symmetry energy $E_{sym,2}$ from the BGBD potential with
$m_n^*>m_p^*$ (left) and for $m_n^*< m_p^*$ (right).}
\label{BGBDe2}
\end{figure}

In Fig.\ \ref{BGBDe2} we compare $E_{sym,2}(\rho)$ and its three
components in the two cases of $m_n^*>m_p^*$ and $m_n^*<m_p^*$. It
is seen that the kinetic and isoscalar contributions are the same
in both cases. However, they have significantly different
isovector potentials $U_{sym,1}$, leading thus to different
$E_{sym,2}(\rho)$ especially at supra-saturation densities.

\begin{figure}[!htb]
\centering
\includegraphics[width=12cm]{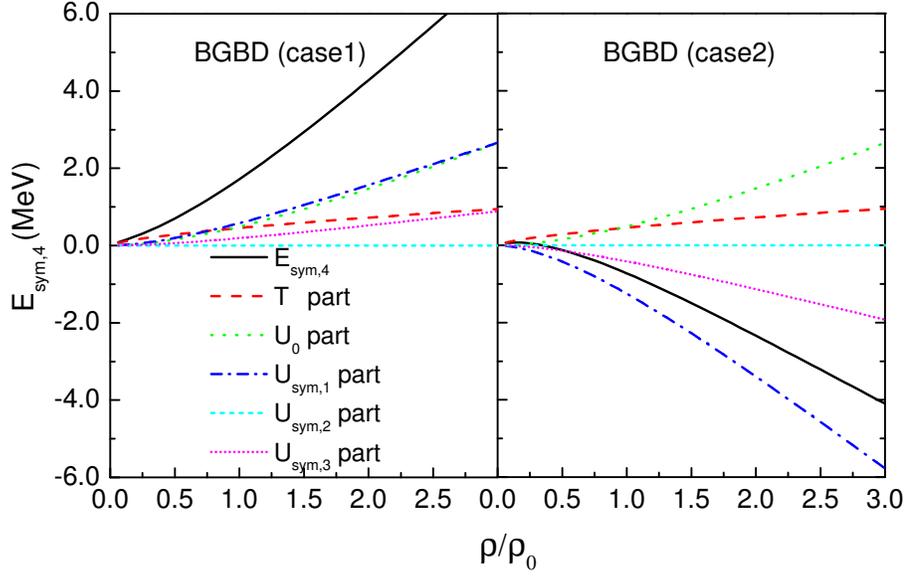}
\caption{The kinetic and various potential contributions to the
fourth-order symmetry energy $E_{sym,4}$ with the BGBD potential
for $m_n^*>m_p^*$ (left) and for $m_n^*< m_p^*$ (right).}
\label{BGBDe4}
\end{figure}

Various contributions to the fourth-order symmetry energies
$E_{sym,4}(\rho)$ in the two cases are compared in Fig.\
\ref{BGBDe4}. Similar to $E_{sym,2}(\rho)$, the contributions of
the $T$ and $U_0$ terms to $E_{sym,4}(\rho)$ are positive and they
are the same in both cases. Interestingly, the $U_{sym,1}$ term
also plays the most important role in determining the high-density
behavior of $E_{sym,4}(\rho)$. It is positive in the case of
$m_n^*>m_p^*$ but negative in the case of $m_n^*<m_p^*$, resulting
in very different behaviors of $E_{sym,4}(\rho)$ at
supra-saturation densities. Moreover, it is interesting to note
that $E_{sym,4}(\rho)$ receives no contribution from the
$U_{sym,2}$ term. This is not surprising because the $U_{sym,2}$
term in the BGBD interaction is momentum independent and its
contribution to $E_{sym,4}(\rho)$ is actually $\frac{1}{12}
\frac{\partial U_{sym,2}(k)}{\partial k}|_{k_F} k_F$=0. On the
contrary, the $U_{sym,3}$ term still contributes to
$E_{sym,4}(\rho)$ via $\frac{1}{4} U_{sym,3}(k_F)$ although it is
momentum independent too. In the two cases considered here, the
contributions from the $U_{sym,3}$ term also have opposite sign.

\begin{figure}[!htb]
\centering
\includegraphics[width=12cm]{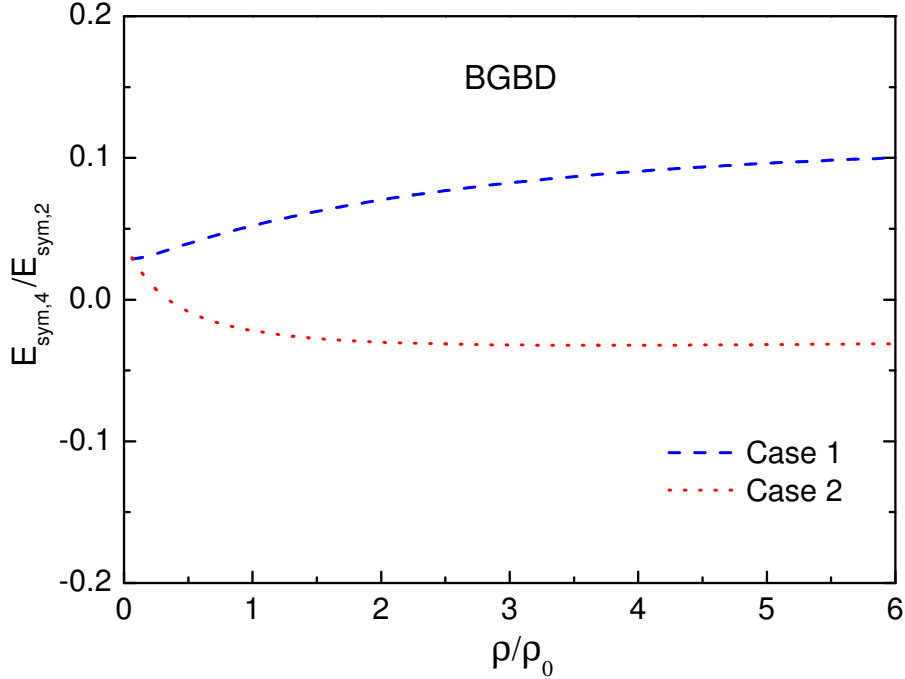}
\caption{The ratio of $E_{sym,4}$ over $E_{sym,2}$ with the BGBD
potential for $m_n^*>m_p^*$ (left) and for $m_n^*< m_p^*$
(right).} \label{BGBDr}
\end{figure}

To compare the fourth-order term $E_{sym,4}(\rho)$ with the
second-order term $E_{sym,2}(\rho)$ more clearly, we show in Fig.\
\ref{BGBDr} their ratio $E_{sym,4}(\rho)/E_{sym,2}(\rho)$ as a
function of the reduced density $\rho/\rho_0$. Obviously, the
relative value of $E_{sym,4}(\rho)$ is generally small. However,
it can reach up to about $\pm 10\%$ at high densities for both
cases of $m_n^*> m_p^*$ and $m_n^*< m_p^*$ . It may thus lead to
an appreciable modification in the proton fraction and therefore
the properties of neutron stars at $\beta$-equilibrium.

\subsection{A modified Gogny Momentum-Dependent-Interaction}\label{MDI}

In this subsection, we discuss the symmetry energy obtained from
the MDI interaction \cite{Das03}, which is derived from the
Hartree-Fock approximation using a modified Gongy effective
interaction \cite{dec}
\begin{eqnarray}\label{mdi}
&&U(\rho,\delta,\vec p,\tau) = A_u(x)\frac{\rho_{\tau'}}{\rho_0}
+A_l(x)\frac{\rho_{\tau}}{\rho_0}\nonumber\\
&&+B(\frac{\rho}{\rho_0})^{\sigma}(1-x\delta^2)-8\tau
x\frac{B}{\sigma+1}\frac{\rho^{\sigma-1}}{\rho_0^{\sigma}}\delta\rho_{\tau'}
\nonumber \\&& +\frac{2C_{\tau,\tau}}{\rho_0} \int
d^3p'\frac{f_{\tau}(\vec r,\vec p')}{1+(\vec p-\vec
p')^2/\Lambda^2}+\frac{2C_{\tau,\tau'}}{\rho_0} \int
d^3p'\frac{f_{\tau'}(\vec r,\vec p')}{1+(\vec p-\vec
p')^2/\Lambda^2}.
\end{eqnarray}
In the above, $\tau=1/2$ ($-1/2$) for neutrons (protons) and
$\tau\neq\tau'$; $\sigma=4/3$ is the density-dependence parameter;
$f_{\tau}(\vec r,\vec p)$ is the phase space distribution function
at coordinate $\vec{r}$ and momentum $\vec{p}$. The parameters $B,
C_{\tau,\tau}, C_{\tau,\tau'}$ and $\Lambda$ are obtained by
fitting the nuclear matter saturation properties \cite{Das03}. The
momentum dependence of the symmetry potential stems from the
different interaction strength parameters $C_{\tau,\tau'}$ and
$C_{\tau,\tau}$ for a nucleon of isospin $\tau$ interacting,
respectively, with unlike and like nucleons in the background
fields. More specifically, $C_{unlike}=-103.4$ MeV while
$C_{like}=-11.7$ MeV. The quantities
$A_{u}(x)=-95.98-x\frac{2B}{\sigma +1}$ and
$A_{l}(x)=-120.57+x\frac{2B}{\sigma +1}$ are parameters. The
parameters $B$ and $\sigma$ in the MDI single-particle potential
are related to the $t_0$ and $\alpha$ in the Gogny effective
interaction via $t_0 = \frac{8}{3} \frac{B}{\sigma+1}
\frac{1}{\rho_0^{\sigma}}$ and $\sigma = \alpha + 1$ \cite{dec}.
The parameter $x$ is related to the spin(isospin)-dependence
parameter $x_0$ via $x=(1+2x_0)/3$ \cite{Xuli10a}. On expanding
the single-nucleon potential in $\delta$, the first four terms are
\begin{eqnarray}
&& U_0(\rho,k) = U_{n/p}|\, _{\delta=0}
\nonumber \\&& = \frac{(A_l + A_u)}{2} \frac{ \rho} {\rho_0} +
B(\frac{\rho}{\rho_0})^{\sigma} +
\frac{2(C_{\tau,\tau}+C_{\tau,\tau'})}{\rho_0} \frac{2}{h^{3}}\pi
\Lambda ^{3} \nonumber  \\&& \times \left[
\frac{p_{F}^{2}+\Lambda^{2}-p^{2}}{2p\Lambda }\ln
\frac{(p+p_{F})^{2} +\Lambda ^{2}}{(p-p_{F})^{2}+\Lambda ^{2}} +
\frac{2p_{F}}{\Lambda }-2\tan ^{-1}\frac{p+p_{f}} {\Lambda }+2\tan
^{-1}\frac{p-p_{f}}{\Lambda }\right],\nonumber\\
\end{eqnarray}
\begin{eqnarray}
&&U_{sym,1}(\rho,k)=  \pm \frac{1}{1!} \frac{\partial
U_{n/p}}{\partial \delta}|\, _{\delta=0}
\nonumber \\&& = \frac{(A_l - A_u)}{2} \frac{ \rho} {\rho_0} - 2
x\frac{B}{\sigma+1}\frac{\rho^{ \sigma}}{\rho_0^{\sigma}} +
\frac{2(C_{\tau,\tau} - C_{\tau,\tau'})}{\rho_0}  \frac{2 p_F^2
\pi \Lambda^2} {3h^3p} \ln \frac{(p+p_{F})^{2} +\Lambda
^{2}}{(p-p_{F})^{2}+\Lambda ^{2}},\nonumber\\
\end{eqnarray}
\begin{eqnarray}
&& U_{sym,2}(\rho,k)= \frac{1}{2!} \frac{\partial^2
U_{n/p}}{\partial \delta^2}|\, _{\delta=0} \nonumber \\&& =  -Bx
(\frac{\rho}{\rho_0})^{\sigma} + \frac{2Bx}{1+\sigma}
(\frac{\rho}{\rho_0})^{\sigma}
\nonumber \\&& +
\frac{(C_{\tau,\tau}+C_{\tau,\tau'})}{3\rho_0} \frac{p_F^2 \pi
\Lambda^2} {9h^3p} \left[ \frac{4 p
\,p_F(p^2-p_F^2+\Lambda^2)}{[(p+p_F)^2+\Lambda^2][(p-p_F)^2+\Lambda^2]}
- \ln \frac{(p+p_{F})^{2} +\Lambda ^{2}}{(p-p_{F})^{2}+\Lambda
^{2}} \right],\nonumber\\
\end{eqnarray}
\begin{eqnarray}
&&U_{sym,3}(\rho,k)=  \pm \frac{1}{3!} \frac{\partial^3
U_{n/p}}{\partial \delta^3}|\, _{\delta=0}
\nonumber\\ && =   -  \frac{(C_{\tau,\tau}-C_{\tau,\tau'})}{3\rho_0}
\frac{4 p_F^2 \pi \Lambda^2} {81h^3p}
\nonumber \\&& \times
 \left[ \frac{2p{p_F}\left(2p^6 -
3{p_F}^6 + 5{p_F}^2{\Lambda}^4 + 2{\Lambda}^6 + p^4\left(-7{p_F}^2
+ 6{\Lambda}^2 \right) + p^2\left(8{p_F}^4 -
2{p_F}^2{\Lambda}^2+6{\Lambda}^4 \right)
\right)}{[(p+p_F)^2+\Lambda^2]^2[(p-p_F)^2+\Lambda^2]^2}
\right.\nonumber \\&& \left.-   \ln \frac{(p+p_{F})^{2} +\Lambda
^{2}}{(p-p_{F})^{2}+\Lambda ^{2}}\right].
\end{eqnarray}

According to Eq.(\ref{Esym2}), the second-order symmetry energy
$E_{sym,2}(\rho)$ is
\begin{eqnarray}
&&E_{sym,2}(\rho) =  \frac{1}{3} t(k_F) + \frac{1}{6}
\frac{\partial U_0}{\partial k}\mid _{k_F}\cdot k_F +
\frac{1}{2}U_{sym,1}(\rho,k_F)
\nonumber \\= && \frac{\hbar^2}{6m}
\left(\frac{3\pi^2}{2}\right)^{2/3} \rho^{2/3}
\nonumber\\ + && \frac{(C_{\tau,\tau}+C_{\tau,\tau'})}{3\rho_0}
\frac{\pi\Lambda^2}{h^3}\left[4p_F-\left(2p_F+\frac{\Lambda^2}{p_F}\right)\textrm{ln}\left(\frac{4p_F^2+\Lambda^2}{\Lambda^2}\right)\right]
\nonumber \\+ && \frac{(A_l - A_u)}{4} \frac{ \rho} {\rho_0} -
x\frac{B}{\sigma+1}\frac{\rho^{ \sigma}}{\rho_0^{\sigma}}
\nonumber \\+ &&
\frac{(C_{\tau,\tau}-C_{\tau,\tau'})}{3\rho_0}
\frac{\pi\Lambda^2}{h^3}2p_F\textrm{ln}\left(\frac{4p_F^2+\Lambda^2}{\Lambda^2}\right),
\end{eqnarray}
and according to Eq.(\ref{Esym4}) the fourth-order symmetry energy
$E_{sym,4}(\rho)$ is
\begin{eqnarray}
&&E_{sym,4}(\rho) = \frac{\hbar^2}{162m}
\left(\frac{3\pi^2}{2}\right)^{2/3} \rho^{2/3}
\nonumber\\ & + &
\left[ \frac{5}{324}\frac{\partial U_0(\rho,k)}{\partial k}|_{k_F}
k_F - \frac{1}{108} \frac{\partial^2 U_0(\rho,k)}{\partial
k^2}|_{k_F} k_F^2 +\frac{1}{648} \frac{\partial^3
U_0(\rho,k)}{\partial k^3}|_{k_F} k_F^3 \right.
\nonumber \\& - & \left.
\frac{1}{36} \frac{ \partial U_{sym,1}(\rho,k)}{\partial k}|_{k_F}
k_F + \frac{1}{72} \frac{ \partial^2 U_{sym,1}(\rho,k)}{\partial
k^2}|_{k_F} k_F^2 + \frac{1}{12} \frac{\partial
U_{sym,2}(\rho,k)}{\partial k}|_{k_F} k_F + \frac{1}{4}
U_{sym,3}(\rho,k_F)\right]
\nonumber \\&=&
 \frac{\hbar ^2}{162m} (\frac{3\pi^2 \rho}{2})^{\frac{2}{3}}
- \frac{C_{\tau,\tau}}{3^{5}\rho _{0}\rho }\left( \frac{4\pi
}{h^{3}}\right) ^{2}\Lambda ^{2}\left[ 7\Lambda ^{2}p_{f}^{2}\ln
\frac{4p_{f}^{2}+\Lambda ^{2}}{\Lambda ^{2}}-\frac{4(7\Lambda
^{4}p_{f}^{4}+42\Lambda
^{2}p_{f}^{6}+40p_{f}^{8}}{(4p_{f}^{2}+\Lambda ^{2})^{2}}\right]
\nonumber \\&-&\frac{C_{\tau,\tau'}}{3^{5}\rho _{0}\rho }\left( \frac{4\pi
}{h^{3}}\right)
^{2}\Lambda ^{2}\left[ (7\Lambda ^{2}p_{f}^{2}+16p_{f}^{4})\ln \frac{%
4p_{f}^{2}+\Lambda ^{2}}{\Lambda
^{2}}-28p_{f}^{4}-\frac{8p_{f}^{6}}{\Lambda ^{2}}\right].\nonumber\\
\end{eqnarray}
As one expects, the above expressions are identical to those derived
directly from the exact MDI EOS using \cite{Chen09}
\begin{eqnarray}
E_{\mathrm{sym},2}(\rho ) &=&\frac{1}{2!}\frac{\partial ^{2}E(\rho
,\delta )}
{\partial \delta ^{2}}|_{\delta =0}\label{Esyme2} \nonumber\\
E_{\mathrm{sym,4}}(\rho ) &=&\frac{1}{4!}\frac{\partial ^{4}E(\rho
,\delta ) }{\partial \delta ^{4}}|_{\delta =0}  \label{Esyme4}.
\end{eqnarray}

\begin{figure}[htb]
\centering
\includegraphics[width=12cm]{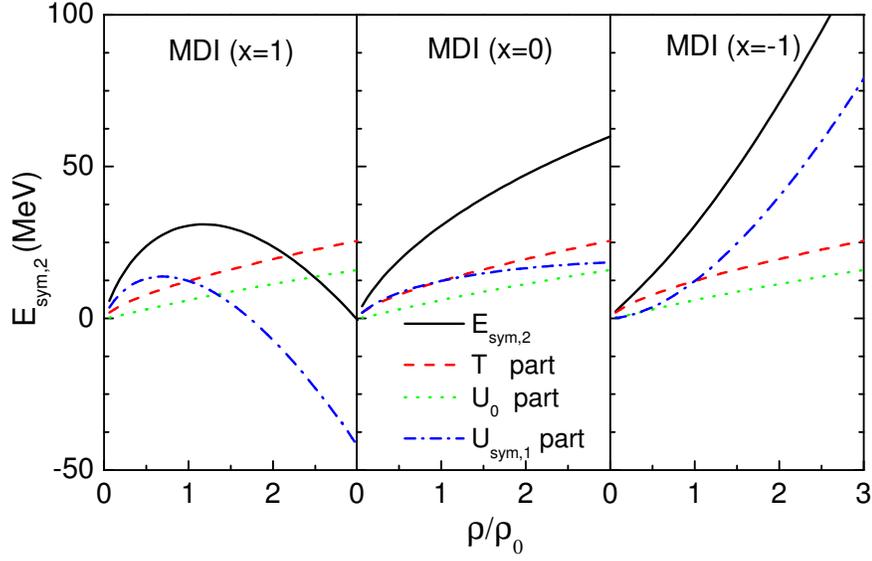}
\caption{The kinetic energy part (T), the isoscalar potential part
($U_0$) and the isovector potential part ($U_{sym,1}$) of the
symmetry energy $E_{sym,2}$ from the MDI interaction with $x=1$, 0
and -1.} \label{MDIe2}
\end{figure}

In Fig.\ \ref{MDIe2}, we show the kinetic (T), isoscalar ($U_0$)
and isovector ($U_{sym,1}$) potential contributions to $E_{sym,2}$
for the three different spin (isospin)-dependence parameter $x=1$,
0, and -1. We notice that the kinetic ($T$) and the isoscalar
potential ($U_0$) contributions are the same for the three
different $x$ values. As pointed out in Ref.\ \cite{Das03}, it is
the isovector potential $U_{sym,1}$ that is causing the different
density dependence of $E_{sym,2}$. For instance, with $x=1$ the
$U_{sym,1}$ term decreases very quickly with increasing density
and thus results in a super-soft symmetry energy at
supra-saturation densities. On the contrary, the symmetry energy
$E_{sym,2}$ at supra-saturation densities is very stiff for both
$x=0$ and $x=-$1 as the contribution of the $U_{sym,1}$ term
becomes very positive with smaller values of $x$.

\begin{figure}[htb]
\centering
\includegraphics[width=12cm]{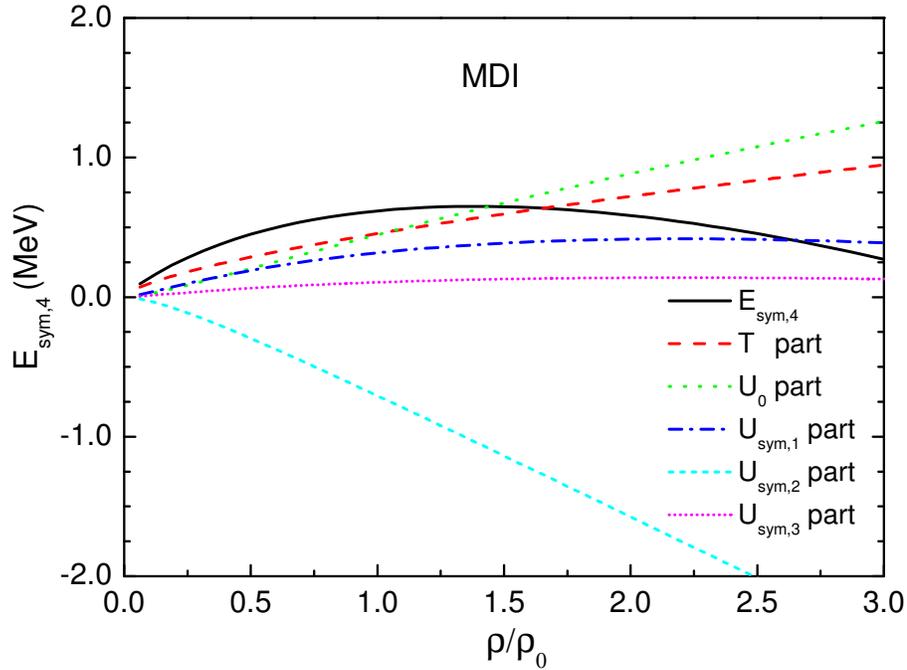}
\caption{The kinetic energy and potential contributions to the
fourth-order symmetry energy $E_{sym,4}$ from the MDI
interaction.} \label{MDIe4}
\end{figure}

Unlike the second-order term $E_{sym,2}$, the fourth-order
symmetry energy $E_{sym,4}$ is independent of the spin
(isospin)-dependence parameter $x$. Shown in Fig.\ \ref{MDIe4} are
the various contributions to the fourth-order symmetry energy
$E_{sym,4}$. Comparing these with the results obtained using the
BGBD in Fig.\ \ref{BGBDe4}, we find that the $T$ and $U_0$ terms
from these two interactions are almost identical. However, there
exists some differences for other terms. For the MDI interaction,
the $U_{sym,2}$ term is negative and becomes very important for
determining $E_{sym,4}$. One the contrary, the contributions from
the $U_{sym,1}$ and $U_{sym,3}$ terms are positive and they are
relatively small as compared to $U_{sym,2}$. Generally, the
behavior of $E_{sym,4}$ from the MDI interaction is very similar
to that from the BGBD interaction for the case of $m_n^*>m_p^*$.

\begin{figure}[htb]
\centering
\includegraphics[width=12cm]{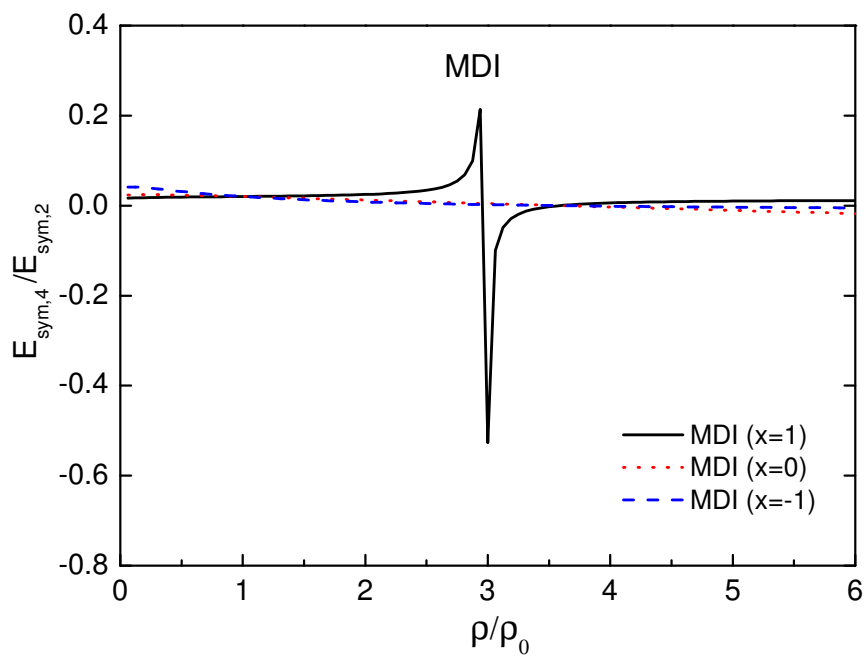}
\caption{The ratio of $E_{sym,4}$ over $E_{sym,2}$ obtained from
the MDI interaction as a function of reduced density $\rho/\rho_0$
for $x=1$, 0, and -1.} \label{MDIr}
\end{figure}

To compare more directly $E_{sym,4}$ with $E_{sym,2}$, their ratio
$E_{sym,4}/E_{sym,2}$ is plotted in Fig.\ \ref{MDIr} as a function
of reduced density for $x=1$, 0, and -1. It is seen that with
$x=1$ there is a sharp break in the curve around $3\rho_0$. This
is because the second-order symmetry energy $E_{sym,2}$ changes
from positive to negative around $3\rho_0$ in this case. However,
this is not the case for both $x=-1$ and $x=0$ where $E_{sym,2}$
remains positive at all densities.  In all cases, $E_{sym,4}$ is
very small compared to $E_{sym,2}$. For both BGBD and MDI
interactions, the small values of $E_{sym,4}$ up to several times
the normal density clearly shows that the parabolic approximation
of the EOS is well justified for most purposes. However, cares
have to be taken in evaluating the core-crust transition density
where the energy curvatures are involved \cite{XCLM09}.

\section{Summary}\label{summary}

In summary, using the Hugenholtz-Van Hove theorem we have derived
general expressions for the quadratic and quartic symmetry
energies in terms of single-particle potentials in isospin
asymmetric nuclear matter. Identical results are obtained by using
two approaches, i.e., one based on the single-particle potential
and the other based on the total energy, that are physically
identical although mathematically different. By using the derived
analytical formulas, the symmetry energies are explicitly
separated into the kinetic and several potential parts. The
formalism is applied to two typical single-nucleon potentials,
namely the Bombaci-Gale-Bertsch-Das Gupta (BGBD) potential and the
modified Gogny Momentum-Dependent-Interaction (MDI), that are
widely used in transport model simulations of heavy-ion reactions.
We find that for both interactions the isovector potential is
responsible for the uncertain high density behavior of the
quadratic symmetry energy. Also, the magnitude of the quartic
symmetry energy in both cases is found to be significantly smaller
than that of the quadratic symmetry energy. We expect that the
analytical formulas for the nuclear symmetry energies derived in
the present study will be useful in extracting reliable
information about the EOS of neutron-rich nuclear matter from
heavy-ion reactions.

\begin{large}
\textbf{Acknowledgements}
\end{large}

This work is supported in part by the US National Science
Foundation grants PHY-0757839 and PHY-0758115, the Research
Corporation under grant No.7123, the Welch Foundation under grant
No. A-1358, the Texas Coordinating Board of Higher Education grant
No.003565-0004-2007, the National Natural Science Foundation of
China grants 10735010, 10775068, 10805026, and 10975097, Shanghai
Rising-Star Program under grant No. 06QA14024, the National Basic
Research Program of China (973 Program) under Contract No.
2007CB815004 and 2010CB833000.

\end{document}